\documentclass[12pt]{iopart}
\usepackage{graphicx}

\begin{document}

\title[]{Effect of ``wandering'' and other features of energy transfer by fast electrons in a direct-drive inertial confinement fusion target}

\author{S~Yu~Gus'kov$^1$, P~A~Kuchugov$^{1,2}$\footnote{Present address: Keldysh Institute of Applied Mathematics of RAS, Miusskaya Sq.~4, Moscow, 125047, Russia}, R~A~Yakhin$^1$ and N~V~Zmitrenko$^2$}

\address{$^1$ P.N.~Lebedev Physical Institute of RAS, Leninskiy Av., 53, Moscow, 119991, Russia}
\address{$^2$ Keldysh Institute of Applied Mathematics of RAS, Miusskaya Sq., 4, Moscow, 125047, Russia}
\ead{pkuchugov@gmail.com}
\vspace{10pt}

\begin{abstract}
The heating of inertial confinement fusion (ICF) target by fast electrons, which are generated as a result of laser interaction with expanding plasma (corona) of a target, is investigated theoretically. It is shown that due to remoteness of the peripheral region, where electrons are accelerated, a significant portion of these particles, moving in corona and repeatedly crossing it due to reflection in a self-consistent electric field, will not hit into the compressed part of target. Using the modern models of fast electron generation, it is shown that in a typical target designed for spark ignition, the fraction of fast electrons that can pass their energy to compressed part of target is enough small. Only 12\% of the total number of fast electrons can do it. Such an effect of ``wandering'' of fast electrons in corona leads to a significant decrease in a negative effect of fast electrons on target compression. Taking into account the wandering effect, the distribution of energy transmitted by fast electrons to different parts of target and the resulting reduction of deuterium-tritium (DT) fuel compression are established.
\end{abstract}

%
\vspace{2pc}
\noindent{\it Keywords}: ICF target, direct-drive, spark ignition, laser-accelerated fast electrons.
%
%
%
%

\maketitle

\section{Introduction\label{sec:001}}

In the problem of spark ignition of a direct-drive thermonuclear target, one of the key questions is a possible preheat of target by suprathermal (fast) electrons. Such particles are generated as a result of the development of plasma instabilities due to laser interaction with an extended plasma of evaporated part of target (corona). The additional (with respect to a shock wave) preheat by fast electrons leads to decreasing target's compression and, therefore, thermonuclear gain reduction. Laser-induced  generation of  fast charged particles is a fundamental phenomenon in high energy density physics. A huge amount of papers is devoted to this subject (see the reviews~\cite{Craxton2015,Kemp2014,Guskov2013} and the book~\cite{Atzeni2004}). Contribution of different  mechanisms of generation, as well as efficiency of laser energy conversion into fast electron energy, energy and angular distributions of laser-accelerated particles depend both on the parameters of laser pulse (intensity, wavelength and duration) and on the characteristics  of plasma in the interaction region (temperature, density and length scale). Other things being equal, the characteristic energy (temperature) of fast electrons increases with increasing the coupling parameter $I \lambda^2$ ($I$ and $\lambda$  are  the laser intensity and wavelength, respectively) which designates the oscillatory energy of electron in a laser field.

Recent experimental and theoretical studies show that in a spherical target designed for spark ignition, when it is irradiated with Nd-laser pulse with an intensity in the range of $10^{14}-10^{15}$~W/cm$^2$, two types of plasma instabilities, in particular, two-plasmon decay (TPD) and stimulated Raman scattering (SRS), are responsible for fast electron generation~\cite{Rosenbluth1972,Liu1974}. Currently, there is no complete physical model of fast electron generation in extended plasma of ICF target. Some papers report about 30-70~keV temperature of the fast electrons  generated in such irradiation conditions. An even greater uncertainty exists in relation to a fraction of laser energy, which is transformed into fast electron energy.

In this regard, the recently published paper~\cite{Rosenberg2018}, which presents the results of experiments under conditions close to irradiation of a direct-drive target at the NIF~\cite{Moses2011}, is of great importance. In these experiments, a flat target was irradiated by the laser pulse of 3rd harmonic of Nd-laser radiation with an energy of 0.2~MJ, duration of about 8~ns, and intensity of $6 \cdot 10^{14}$ -- $1.6 \cdot 10^{15}$~W/cm$^2$. The main results, obtained on the base of comparison of the data from various measurements and numerical calculations, are as follows. The generation of fast electrons is mainly attributed to SRS and occurs in the plasma region with quarter-critical density. The temperature of fast electrons was determined as about 50~keV, while the energy of fast electron emission -- about 1\% of laser energy. Despite the fact that this study is based on using a flat target, its results can be viewed as the first set of experimental data relating, simultaneously, to the origin of fast electron generation and their emission parameters. The main argument of this conclusion is the interaction of a real laser pulse intended for direct-drive ICF target with extended plasma.

In the present paper the features of energy transfer by fast electrons in a direct-drive ICF target are investigated using the data from~\cite{Rosenberg2018} as a scales of the parameters of fast electron emission. In the second section the conditions for the energy transfer by fast electrons in a typical ICF target, designed for spark ignition with an absorbed laser energy of 1.5~MJ, are discussed. The characteristics of fast electron emission as well as the temporal evolution of target state during implosion, used in this paper, are represented. In the third section, the features of interaction of fast electrons with the different parts of target, namely, with expanding corona and compressed part of target are studied. In the fourth section, a quantitative model of fast electron energy deposition in the different parts of ICF target is developed. The dependence of the energy that fast electrons transfer to thermonuclear fuel, as well as the dependence of reduction of its final density on fast electron emission parameters are determined.

\section{Problem statement: dynamics of target's implosion and characteristics  of  fast electron emission\label{sec:002}}

Direct-drive ICF targets intended for modern ignition experiments are double-shell ones with an outer layer, ablator, consisting of  light element material (such as polystyrene) and inner layer of DT-ice. The targets are designed  for absorbed energy 1.5-1.7~MJ of  the 3rd or 2nd harmonic of Nd-laser radiation. The target parameters meet the conditions for improved stability of implosion with profiled laser pulse of about 400~TW peak power at contrast not exceeding 40. The most important factors of improved stability are relatively low aspect ratio (ratio of shell radius to its thickness) and large fraction of  evaporated ablator mass, which contributes to the ablative stabilization~\cite{Takabe1985} of hydrodynamic instabilities. The adjustment of the parameters of such a target and laser pulse ensures achievement of maximal shell velocity up to the values of 400~km/s and evaporation of the most part of ablator (75-90\% of its mass) at the DT-shell aspect ratio not exceeding 5-10.

\begin{figure}[!ht]
  \centering
  \includegraphics[width=0.6\textwidth]{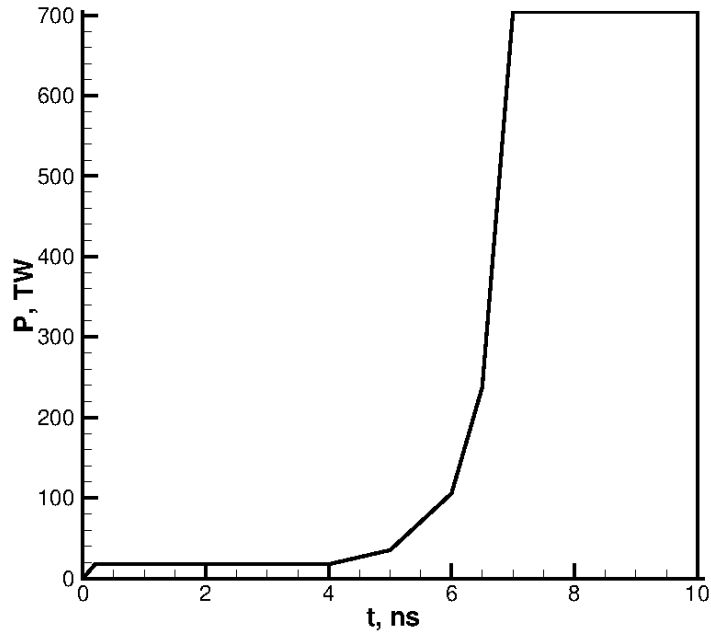}
  \caption{The dependence of the power of the laser pulse related to the absorbed energy, the fraction of which is 0.57.\label{fig:001}}
\end{figure}

The initial parameters and implosion characteristics of this kind of targets, discussed in various laboratories~\cite{Craxton2015,Brandon2014,Belkov2015}, differ slightly from each other. Therefore, the features of the kinetics of fast electrons in such targets have a general nature. These features are discussed in this paper in relation  to the baseline target~\cite{Belkov2015} designed for irradiation conditions expected for the megajoule laser facility of the Russian project~\cite{Garanin2011}. These conditions suggest the action of a profiled pulse with the energy 2.6~MJ of the 2nd harmonic of Nd-laser radiation. The outer radius of the target is $R = 1597$~$\mu$m. The initial thicknesses of CH-ablator and DT-ice layer are $\Delta_a = 34$~$\mu$m and $\Delta_{s} = 149$~$\mu$m, respectively, and their masses are $M_a = 1.12$~mg and $M_{s} = 1.06$~mg. The aspect ratio of DT-ice layer is $\left(R - R_a\right) / \Delta_{s} = 10.5$. The shell is filled with DT-gas with density of $10^{-3}$~g/cm$^3$. The target is designed for absorbed energy 1.5~MJ of the 2nd harmonic of Nd-laser radiation. The time dependence of the pulse power related to the laser energy absorbed in the target is shown on Fig.~\ref{fig:001}. The absorption coefficient is 0.57 and was calculated using a combined model of ray-tracing and wave optics in the framework of the RAPID code~\cite{Belkov2015,Rozanov1985}. The code  provides a joint solution of one-dimensional (1D) hydrodynamic equations and Maxwell equations (near the turning point of ray) with taking into account inverse bremsstrahlung and resonant absorption mechanisms. On account of the considerable length of target corona, the role of a resonant absorption is insignificant. The fraction of laser energy absorbed due to this mechanism is less than 0.1\%. At the same time, the calculations show a strong effect of laser light refraction in extended corona's plasma, which leads to the fact that the region with the maximum energy release shifts from the region with the critical density to a less dense peripheral region~\cite{Belkov2015}.

A pulse with a duration of 10~ns has a power of 10~TW in its initial part, up to 4~ns. Then a power increases to a value of 400~TW from 4 to 7~ns, after that it doesn't change until the pulse terminates (see Fig.~\ref{fig:001}). The gain of the considered target, without taking into account its heating by fast electrons, is 21~\cite{Belkov2015}. It was obtained in the calculation using numerical code DIANA~\cite{Belkov2015,Zmitrenko1983}, which provides the solution of the equations of 1D two-temperature hydrodynamics with energy sources due to inverse bremsstrahlung absorption of laser radiation and $\alpha$-particle heating, as well as with taking into account all main relaxation and transport processes in plasma and real equation of state. Exactly the data of this calculation will be used to characterize the state of target, in which fast electrons are decelerated.

In relation to fast electron interaction with target, only the time period $\tau_h$ from $t_0 = 7$~ns to $\tau_L = 10$~ns, which refers to high intensity part of the pulse, is relevant. Based on the data of~\cite{Rosenberg2018}, it is assumed that fast electrons are generated in the time period $\tau_h$ near the surface with quarter-critical density $n_{qc} = 0.25 n_c$ ($n_c = 10^{21} \lambda^{-2}$~cm$^{-3}$, where the laser radiation wavelength is measured in $\mu$m). Further, to be short, this surface will be called as the $n_{qc}$-surface. It is assumed that fast electrons are generated isotropically. The main prerequisite for this is a strong effect of laser light refraction in an extended corona of ICF target. The already mentioned RAPID-code simulations~{\cite{Belkov2015}} have been performed for the geometry of irradiation of considered target by 192 Gaussian laser beams with an aperture $D$ of 40~cm at a focal length of $F = 660$~cm. The beam parameter $s$ (ratio of the beam radius to the initial radius of target) varied from 0.5 to 1.3. A characteristic result for the case of $s = 1$ is that, depending on the angle $\theta$ between a beam optical axis and radius-vector of a point on target's surface the density in a plasma region corresponding to turning of a beam varies from $0.05 \rho_c$ at $\theta = 90^{\circ}$ to $0.8 \rho_c$  at $\theta = 0^{\circ}$. This means that the refraction will lead to the significant broadening an angular distribution of fast electrons generated near the region with a quarter-critical density, due to the fact of different incidence angles of beam's rays in the region of particle generation. An additional argument for the isotropic angular distribution is the conclusion of experiments~{\cite{Yaakobi2013}} on the large divergence of fast electrons generated due to the TPD.

The generation rate $q_0$ is considered to be constant in accordance with the constant power of the high-intensity part of laser pulse. In addition, to simplify the problem statement, the monoenergetic fast electrons are considered. The ranges of possible values of the initial energy of fast electron $E_0$ and the fraction of laser energy contained in fast electron emission, are chosen around the values recommended in~\cite{Rosenberg2018}, as follows
\begin{equation}\label{eq:001}
  30~\mbox{keV} \leq E_0 \leq 70~\mbox{keV},
\end{equation}
\begin{equation}\label{eq:002}
  0.005 \leq \eta \leq 0.015,
\end{equation}
where $\eta = q_0 \tau_h E_0 / E_L$, $E_L$ is an energy of high-intensity part of the pulse, $E_L = 2.1$~MJ.
The upper limits  in~(\ref{eq:001}) and~(\ref{eq:002}) take into account the possibility of increasing the values $E_0$ and $\eta$ compared with those recommended in~\cite{Rosenberg2018} when 2nd harmonic pulse is used. The lower ones reflect the possibility of reducing these values due to the fact that the length of spherically exploded corona is less than in a flat case. Using an approximation of the monoenergetic spectrum, an analytical model of energy transfer by fast electrons is developing in an inhomogeneous plasma, the various regions of which have strongly different stopping powers. This model is using to analyze the energy transfer by electrons with an initial energy varying in a wide range of values~({\ref{eq:001}}). Taken together, such a techniques, to a certain extent, is a multi-group approach by energy spectrum.

\begin{figure}[!ht]
  \centering
  \includegraphics[width=0.6\textwidth]{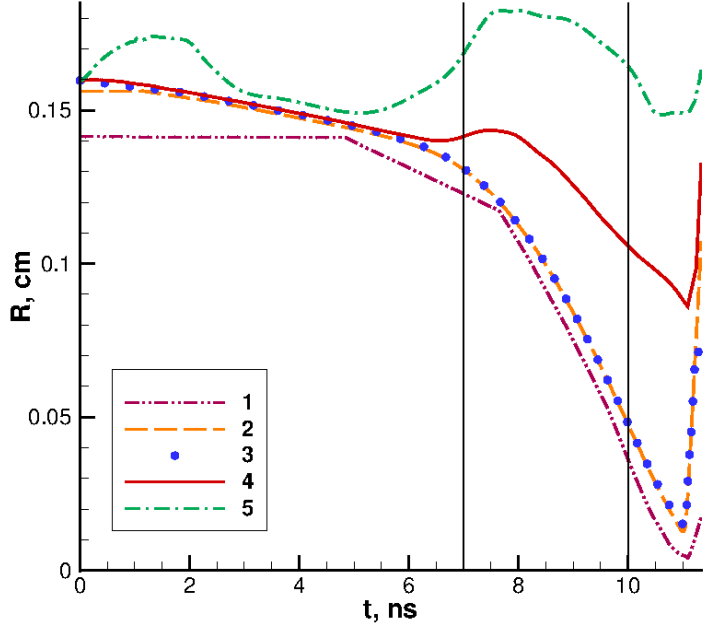}
  \caption{Time dependencies of the surface positions of the various regions of target. Curve 1 corresponds to the inner surface of DT-shell, 2 -- to the surface between DT-fuel and ablator, 3 -- to the evaporation surface (the outer surface of a non-evaporated part of ablator), 4 -- to a surface with critical density, 5 -- to a surface with a quarter-critical density. The vertical lines denote the time interval corresponding to the high-intensity part of pulse, during which the generation of fast electrons occurs.\label{fig:002}}
\end{figure}

Below the features of shell implosion during the period of fast electron generation $t_0 \leq t \leq \tau_L$ are considered. At the time moment $t_0 = 7$~ns, with the beginning of the high-intensity part of laser pulse, fast convergence of the shell to the center begins. At this moment the fraction of evaporated ablator mass is 25\%, the average density of non-evaporated ablator and average density of DT-layer are 8.83~g/cm$^3$ and 0.66~g/cm$^3$, respectively, and the thicknesses of these layers are 2.4~$\mu$m and 74~$\mu$m. The thicknesses of the layers are significantly less than the radius of the shell, which is about 1350~$\mu$m in this moment. On Fig.~\ref{fig:002} the $R-t$ diagrams for  surfaces of the various areas in expanding corona and compressed shell are shown. First of all, it should be noted the features of temporal evolution during the time interval $\tau_h = \tau_L - t_0 = 3$~ns of the positions of two surfaces, namely, the $n_{qc}$-surface, where fast electrons are generated (curve 5 on Fig.~\ref{fig:002}), and the outer surface of compressed shell that is the evaporation surface (further the $n_s$-surface -- curve 3 on Fig.~\ref{fig:002}). Firstly, the radius of the $n_{qc}$-surface remains approximately constant around the value $R_{qc} = 1850$~$\mu$m. Secondly, the radius of the $n_s$-surface with an initial value $R_{s0} = 1350$~$\mu$m (at $t_0 = 7$~ns) decreases with a constant implosion velocity $V_i = 350$~km/s.

Thus, fast electrons are generated near the surface, the radius of which can be considered constant throughout the entire period of particle generation. In turn, the radius of compressed shell, constantly decreases in the same time period. From this it follows that, under the assumption of an isotropy of fast electron emission, only a part of the produced particles can get into the compressed shell and to heat it, while other particles will move past it. Moreover, the fraction of the heating particles is constantly decreasing as a target is imploded.

\section{Effect of wandering and fraction of the heating fast electrons\label{sec:003}}

Let us determine the fraction of heating fast electrons that are generated on the $n_{qc}$-surface and in the course of their further movement have an opportunity to get into compressed  shell, i.e. to cross the $n_s$-surface.

\begin{figure}[!ht]
  \centering
  \includegraphics[width=0.6\textwidth]{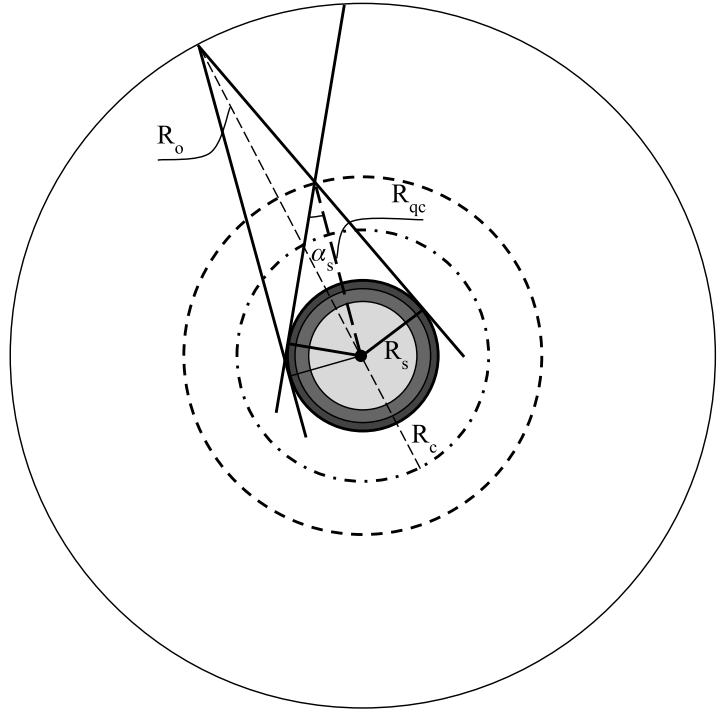}
  \caption{The trajectories of the motion of fast electrons generated on the $n_{qc}$-surface (dotted circle). The dash-dotted circle denotes the surface with  critical density. The two central spherical layers denote non-evaporated ablator and DT-layer. The outer circumference denotes the surface of reflection of fast electrons inside the target due to a self-consistent electric field.\label{fig:003}}
\end{figure}

From Fig.~\ref{fig:003}, which illustrates motion of fast electrons, it is clear that only the fraction of fast electrons generated in the direction toward the center of target inside the solid angle $\Delta \Omega_s$ and ones generated in the direction from the center inside the vertical solid angle can get inside the compressed part of target. The rest of the particles, ``wandering electrons'', which are generated outside these angles, will move in the corona, not hitting into compressed shell. Moreover, the fraction of heating fast electrons does not depend on the radius of the surface, on which the particles born in the outward direction are reflected inside the target due to a self-consistent electric field.

\begin{figure}[!ht]
  \centering
  \includegraphics[width=0.6\textwidth]{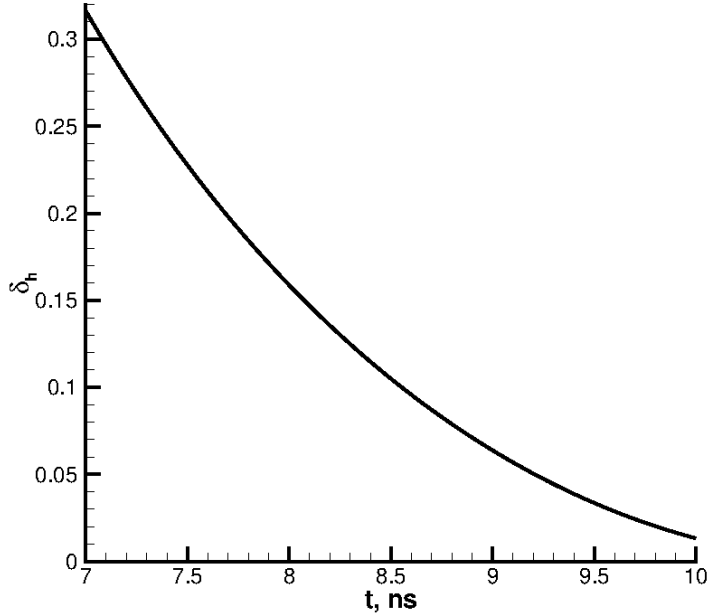}
  \caption{Time dependence of the ratio of the power of the source of heating fast electrons to the power of the source of all fast electrons.\label{fig:004}}
\end{figure}

At a constant power $q_0$ of fast electron source, the total number of these particles is $N_0 = q_0 \tau_h$. Then the power of the source of heating fast electrons is given by
\begin{equation}\label{eq:003}
  q_h = \delta_h q_0,
\end{equation}
where
\begin{equation}\label{eq:004}
  \delta_h = \frac{2 \Delta \Omega_s}{4 \pi} = 1 - \cos{\alpha_s},
\end{equation}
$\displaystyle \cos{\alpha_s} = \sqrt{1 - \left(R_s / R_{qc}\right)^2}$, $R_s = R_{s0} - V_i \left(t - t_0\right)$, $t_0 \leq t \leq \tau_L$, $\alpha_s$ is angle between the radius of the $n_{qc}$-surface drawn to the point where fast electron is generated and a tangent drawn from the same point to the circle corresponding to the $n_s$-surface. On Fig.~\ref{fig:004} the time dependence of the ratio $\delta_h = q_h / q_0$ for the above parameters of the problem is shown. The initial radius of $n_s$-surface at time $t_0 = 7$~ns is $R_{s0} = 1350$~$\mu$m, the radius of $n_{qc}$-surface is $R_{qc} = 1850$~$\mu$m, which is approximately considered to be unchanged with time, and the implosion velocity is 350~km/s. The ratio $\delta_h$ decreases as target is imploded from about 31\% to 1.5\%. The magnitude
\begin{equation}\label{eq:005}
  \eta(t) = \tau_L^{-1} \int_{t_0}^{t}{\delta_h(t) dt}
\end{equation}
is the fraction of heating fast electrons of the total number of fast electrons at the time $t$, while
\begin{equation}\label{eq:006}
  \eta_h = \tau_L^{-1} \int_{t_0}^{\tau_L}{\delta_h(t) dt} = 1 + \frac{R_{qc}}{2 V_i \tau_h} \left.\left[\xi \sqrt{1 - \xi^2} + \arcsin{\xi}\right]\right|_{\xi_1}^{\xi_2},
\end{equation}
where $\displaystyle \xi_1 = R_{s0} / R_{qc}$, $\displaystyle \xi_2 = \xi_1 - V_i \tau_h / R_{qc}$, gives the fraction of all heating fast electrons of the total number of fast electrons (in the time $\tau_L$). For the considered target it equals about 12\%.

It should be noted that the result obtained does not depend on the type of spectrum of fast electrons, unless it contains too low energy particles that are completely slowed down while moving to the $n_s$-surface. In the latter case, the value of $\delta_h$ is even smaller than the calculation presented above. Moreover, the value of $\delta_h$ is not affected by the scattering of fast electrons, which accompany their deceleration in Coulomb collisions. Finally, the fraction of fast electrons which can transfer their energy to the compressed shell decreases with removing the region of particles generation from the $n_{qc}$-surface to a peripheral region of corona.

\section{Energy transmission from fast electrons\label{sec:004}}

In this section the question of energy transfer by the heating fast electrons to different parts of target is considered in order to determine the fraction of energy that they transmit to DT-fuel. In this case, it is taken into account that the thickness of non-evaporated ablator during the time period $\tau_h$ is significantly less (approximately 600 times) than the radius of the evaporation boundary (see the previous section). On this reason , the solid angle $\Delta \Omega_s$ related to the evaporation surface is very slightly different from the solid angle related to the boundary between ablator and DT-layer. For this reason, the problem is considered in the approximation of one number of heating fast electrons directed to the evaporation boundary and the boundary between ablator and DT-layer.

To calculate a mean free path of fast electron with initial energy $E_0$ in the plasmas of CH-ablator and DT-fuel while its slowing down in Coulomb collisions with plasma electrons and scattering by plasma ions, the results of~\cite{Ribeyre2013} are used, where corresponding expressions are given in a simple form. The mass range of a fast electron $\mu$, which is calculated as a product of plasma density and mean free path, grows as the square of initial particle energy and is inversely proportional to the average ions charge $Z$. In addition, it weakly depends, logarithmically, on the density and temperature of plasma. The hydrodynamic states of corona and compressed shell are strongly different, and therefore the mass ranges in these regions also differ greatly. The characteristic corona's density corresponds to critical plasma density $\rho_c \approx 1.83 \cdot 10^{-3} A / Z \left(\lambda [\mu \mbox{m}] \right)^{-2}$~g/cm$^3$, the value of which for a fully ionized CH-plasma and the 2nd  harmonic radiation of a Nd-laser is about $1.2 \cdot 10^{-2}$~g/cm$^3$. The temperature in the region of critical density is close to 2-3~keV. Accelerated inward shell until the end of laser pulse and, consequently, until the end of fast electrons generation has the density about 10~g/cm$^3$ and the temperature about 10-20~eV, which is determined by shock wave heating. Below the values of fast electron mass ranges in DT-fuel (fully ionized plasma, $Z = 1$) $\mu_{1}$, compressed CH-ablator (degree of ionization $Z = 1$) $\mu_2$ and CH-corona (fully ionized plasma, $Z = 3.5$) $\mu_3$ are, respectively, represented. The Coulomb logarithms were calculated at the critical density and the temperature of 2~keV for CH-corona, at the density of 10~g/cm$^3$ and the temperature of 20~eV for compressed CH-ablator and DT-fuel. The mass ranges are:
\begin{equation}\label{eq:009}
  \mu_1 = 3.47 \cdot 10^{-3} \left(\frac{E}{50}\right)^2,
\end{equation}
\begin{equation}\label{eq:008}
  \mu_2 = 1.1 \cdot 10^{-2} \left(\frac{E}{50}\right)^2,
\end{equation}
\begin{equation}\label{eq:007}
  \mu_3 = 2.1 \cdot 10^{-3} \left(\frac{E}{50}\right)^2,
\end{equation}
where $E_0$ is measured in keV here and further.

\begin{figure}[!ht]
  \centering
  \includegraphics[width=0.6\textwidth]{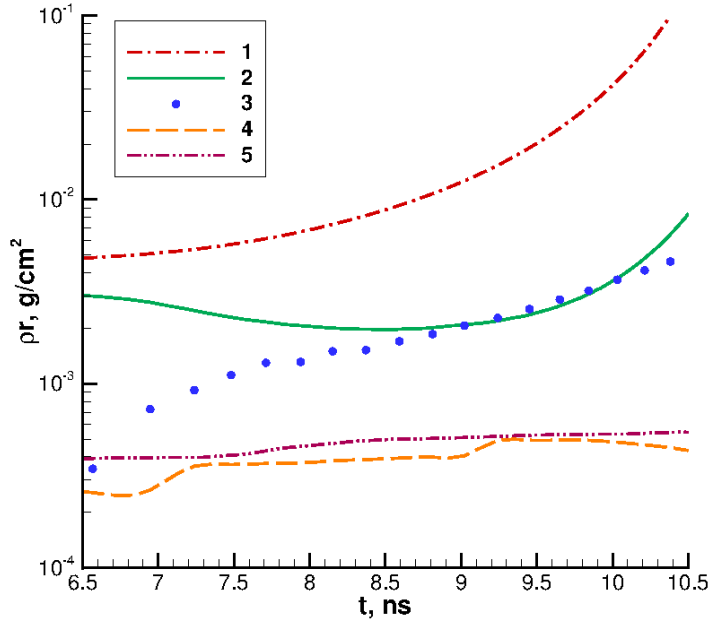}
  \caption{The dependencies of the areal density of various areas on time: 1 -- DT, 2 -- region from the outer boundary of the DT-shell to the evaporation surface (layer of the non-evaporated part of ablator), 3 -- from the evaporation surface to the critical surface, 4 -- from the critical surface to the quarter-critical surface, 5 -- from the quarter-critical surface to the outer edge of corona.\label{fig:005}}
\end{figure}

The calculation of distribution of the energy, which heating fast electrons transfer to the different parts of target, is carried out with use the slowing down model in a layered medium with time-dependent values of areal density in the different layers according to the expression~\cite{Sivukhin1966}
\begin{equation}\label{eq:025}
  \frac{dE}{dr} = -\frac{1}{2E} F,
\end{equation}
where $F \propto A / Z \rho$ is calculated in accordance with~\cite{Ribeyre2013}.
Since the solid angle, in which the heating fast electrons move, is small, their mass ranges in spherical layers can be approximately compared with the areal densities calculated along a radius $\displaystyle \left<\rho r\right> = \int{\rho(r) dr}$. On Fig.~\ref{fig:005} the time dependencies (at $t \geq t_0 = 7$~ns) are shown of the areal densities of various target areas, namely, from the corona expansion surface to the $n_{qc}$-surface (region 5), from the $n_{qc}$-surface to the surface of critical density (hereinafter $n_c$-surface, region 4), from the $n_c$-surface to $n_s$-surface (region 3), from $n_s$-surface to outer boundary of DT-shell (region 2) and from outer boundary of DT-shell to the center of the target (region 1).
Two important circumstances should be noted that make it possible to simplify model of energy transfer. Firstly, the energy transfer to considered target from fast electrons with initial energy in the range~(\ref{eq:001}) occurs during their single pass through the target. This follows from the fact that the fast electron hitting into DT-layer after the first pass through the target is completely inhibited (thermalized). Indeed, the minimum value of DT areal density (at time $t_0 = 7$~ns) is, according to Fig.~\ref{fig:005} (curve 1), $\left<\rho r\right>_1^{\min} = 5 \cdot 10^{-3}$~g/cm$^2$. The fast electron got into DT has energy lower than the initial value $E_0$, because before that it was slowing down in the peripheral parts of target with respect to DT-layer. But even for the electron with initial energy equal to the upper value of the range~(\ref{eq:001}), $E_0 = 70$~keV, the mass range, according to~(\ref{eq:009}), is $\mu_1^{\max} = 6.8 \cdot 10^{-3}$~g/cm$^2$, which is less than twice the minimum possible DT areal density. This means that even the electrons with this maximum possible energy partially decelerated in DT on the first pass, moving to the center, completely slowing down in this layer on the second pass, moving from the center. Secondly, according to Fig.~\ref{fig:005} (curve 4), the areal density in the region from the $n_{qc}$-surface to the $n_{c}$-surface during the period of fast electron generation varies slightly with time and is about $3 \cdot 10^{-4}$~g/cm$^2$. This value is significantly less than the fast electron mass range, which, according to~(\ref{eq:007}), is $4.1 \cdot 10^{-3}$~g/cm$^2$ for the initial energy $E_0 = 70$~keV, and $7.5 \cdot 10^{-4}$~g/cm$^2$ for $E_0 = 30$~keV.

Therefore, it can be assumed that the fast electrons generated in the direction from the center of target and, after reflection in a self-consistent electric field, got into the region of compressed shell, almost do not lose their energy, moving in the corona region with a density lower than the critical one. It means that all heating fast electrons that hit into the compressed shell -- both those that are born in directions away from the center of target, and those that are born towards the center, are effectively slowed down only in the region with density larger than the critical one, having its initial energy on the $n_c$-surface.

Thus, the task is simplified and reduces to finding those parts of the energy which fast electrons transfer to two regions of ablator, namely, to the area of a relatively dense evaporated part of ablator from a surface with  the critical density to the evaporation surface (region 3, according to numbering used on Fig.~\ref{fig:005}) and to the area of non-evaporated ablator (region 2). The rest of fast electron energy is absorbed in DT. According to Fig.~\ref{fig:005}, the areal density of the region 3 increases rapidly with time due to the influx of evaporated mass. The temporal dependence of areal density $\left<\rho r\right>_3$ in this region can be approximated by an exponential function
\begin{equation}\label{eq:010}
  \left<\rho r\right>_3 = 7 \cdot 10^{-4} \exp{\left[1.75 \frac{t - t_0}{\tau_h}\right]},\; t_0 \leq t \leq \tau_L.
\end{equation}
In turn, the time dependence of the areal density of non-evaporated ablator is a slowly varying, non-monotonic function. From 7~ns to 9~ns, the areal density decreases due to decreasing the mass of this region since evaporation of ablator is a stronger effect than its compression. The increase in areal density begins after 9~ns, when the effect of compression becomes dominant. In general, with sufficient accuracy, it can be assumed that the areal density of the region 2 during the entire period of fast electron generation remains constant
\begin{equation}\label{eq:011}
  \left<\rho r\right>_2 = 2.5 \cdot 10^{-3}~\mbox{g/cm}^2,\; t_0 \leq t \leq \tau_L.
\end{equation}
Even the simplest analysis of a fast electron deceleration in the region 3 shows that the range~(\ref{eq:001}) of their initial energy corresponds to the different regimes of energy transfer. Indeed, the mass range of fast electron with a minimum energy of 30~keV, according to~(\ref{eq:007}), in the region 3 is $7.6 \cdot 10^{-4}$~g/cm$^2$. The minimum areal density of this region at time $t = 7$~ns is $\left<\rho r\right>_3^{\min} = 7.5 \cdot 10^{-4}$~g/cm$^2$. This means that fast electrons with such an energy almost will not heat a compressed part of the target, and will not have a negative effect on compression. On the contrary, energy transfer by fast electrons of such an energy to the corona's region with subsonic expansion, will increase an ablation pressure and, consequently, will positively effect on compression. Further, the mass range of fast electron with a maximum initial energy of 70~keV, according to~(\ref{eq:007}), is $4.1 \cdot 10^{-3}$~g/cm$^2$. This value is greater than the maximum value of areal density in the region 3, which  reaches the magnitude of $\left<\rho r\right>_3^{\max} = 4 \cdot 10^{-3}$~g/cm$^2$ at the end of the laser pulse. This means that almost all heating fast electrons with the maximum initial energy from range~(\ref{eq:001}) after partial slowing down in the region 3 will hit into the compressed shell. Thus, the character of energy transfer strongly depends on the initial energy of  fast electrons from the range~(\ref{eq:001}).

The quantitative model of energy transferred from fast electrons, depending on their initial energy, is constructed as follows. As a result of the slowing down of a fast electron in a layer with areal density $\left<\rho r\right>$, its initial energy $E_0$, according to~(\ref{eq:025}) decreases to the value
\begin{equation}\label{eq:012}
  E = E_0 \left[1 - \frac{\left<\rho r\right>}{\mu_0}\right]^{1/2},
\end{equation}
where $\mu_0$ is the range of fast electron at decreasing its energy from the initial value $E_0$ to 0.

Then, the energy of fast electron, generated at time $t$, after its deceleration in the region 3 is
\begin{equation}\label{eq:013}
  E_3(t) = E_0 \left[1 - \frac{\left<\rho r\right>_3\left(t\right)}{\mu_3\left(E_0\right)}\right]^{1/2},
\end{equation}
where $t_1$ is the moment, which is determined by the condition of equality of fast electron mass range and areal density in the region 3, which increases with time
\begin{equation}\label{eq:014}
  \mu_3\left(E_0\right) = \left<\rho r\right>_3\left(t\right),
\end{equation}
and the functions $\mu_3 \left(E_0\right)$ and $\left<\rho r\right>_3 \left(t\right)$, used in~(\ref{eq:013}) and (\ref{eq:014}), are given by expressions~(\ref{eq:007}) and~(\ref{eq:010}), respectively.

The fast electron with energy $E_3\left(t\right)$, which was generated at the moment $t$ and then slowed down in the region 3, hit into the region of dense ablator. After slowing down in the region 2, the energy of  fast electron decreases to the value
\begin{equation}\label{eq:015}
  E_2 = E_3(t) \left(1 - \frac{\left<\rho r\right>_2}{\mu_2\left(E_3(t)\right)}\right)^{1/2},\; t_0 \leq t \leq t_2,\; t_2 \leq t_1,
\end{equation}
where the energy $E_3\left(t\right)$ is given by the expression~(\ref{eq:013}); the value of areal density $\left<\rho r\right>_2 = 2.5 \cdot 10^{-3}$~g/cm$^2$ does not change in time (see~(\ref{eq:011})); the mass range of fast electron in the region 2 after its partial slowing down in the region 3, according to~(\ref{eq:008}), is given by
\begin{equation}\label{eq:016}
  \mu_2\left(E_3(t)\right) = 1.1 \cdot 10^{-2} \left(\frac{E_0}{50}\right)^2 \left[1 - \frac{\left<\rho r\right>_1(t)}{\mu_3\left(E_0\right)}\right].
\end{equation}
The time $t_2$ is determined from the condition of equality of  mass range of fast electron with the initial energy $E_3\left(t\right)$ and areal density of the region 2, which, as mentioned above, can be considered as constant
\begin{equation}\label{eq:017}
  \mu_2\left(E_3(t)\right) = \left<\rho r\right>_2 \equiv 2.5 \cdot 10^{-3}~\mbox{g/cm}^2.
\end{equation}
Expression~(\ref{eq:017}), written at the maximum energy of fast electrons got into the region 2. These particles were slowed down with a minimum surface density of the region 3 $\left<\rho r\right>_3^{\min} = 7 \cdot 10^{-4}$~g/cm$^2$. So, expression~(\ref{eq:017}) determines the initial energy of fast electrons, at which they will not hit into the DT-layer:
\begin{equation}\label{eq:018}
  1.1 \cdot 10^{-2} \left(\frac{E_0}{50}\right)^2 \left[1 - \frac{\left<\rho r\right>_3^{\min}}{\mu_3\left(E_0\right)}\right] = \left<\rho r\right>_2.
\end{equation}
Substituting $\left<\rho r\right>_3^{\min} = 7 \cdot 10^{-4}$~g/cm$^2$, $\left<\rho r\right>_2 = 2.5 \cdot 10^{-3}$~g/cm$^2$ and mass range $\mu_3 \left(E_0\right)$, according to~(\ref{eq:007}), into~(\ref{eq:018}), one can obtain, that fast electrons with initial energy less than 37~keV will not get into the DT-layer.

At a constant power of fast electron source, the number of particles that hit into the region 2 and the number of particles that hit into the DT-layer are calculated using integrals
\begin{equation}\label{eq:019}
  N_2 = N_0 \eta_h \eta_h^{-1} \tau_h^{-1} \int_{t_0}^{t_1}{\delta_h(t)dt},
\end{equation}
and
\begin{equation}\label{eq:019b}
  N_1 = N_0 \eta_h \eta_h^{-1} \tau_h^{-1} \int_{t_0}^{t_2}{\delta_h(t)dt},
\end{equation}
respectively.

The energy that fast electrons leave in the region 3 is:
\begin{eqnarray}\label{eq:020}
  \Delta E_3 &=& E_0 \left\{\int_{t_0}^{t_1}{q_h(t) \left\{1 - \left[1 - \frac{\left<\rho r\right>_3(t)}{\mu_3\left(E_0\right)}\right]^{1/2}\right\} dt} + \int_{t_1}^{\tau_L}{q_h(t) dt}\right\} = \nonumber
  \\
  &=& E_0 N_0 \eta_h \left\{1 - \eta_h^{-1} \tau_h^{-1} \int_{t_0}^{t_1}{\delta_h(t) \left[1 - \frac{\left<\rho r\right>_3(t)}{\mu_3\left(E_0\right)}\right]^{1/2} dt}\right\},
\end{eqnarray}
where the function $\delta_h(t)$ is given by expression~(\ref{eq:004}), and the fraction of heating fast electrons $\eta_h$ for the considered target is 0.12 (see~(\ref{eq:006})); $E_0 N_0 \eta_h$ is the total energy of heating fast electrons. The energy which fast electrons leave in the region 2 is:
\begin{eqnarray}\label{eq:021}
  \Delta E_2 &=& E_0 \int_{t_0}^{t_2}{q_h(t) \left[1 - \frac{\left<\rho r\right>_3}{\mu_3\left(E_0\right)}\right]^{1/2} \left\{1 - \left(1 - \frac{\left<\rho r\right>_2}{\mu_2\left(E_1\right)}\right)^{1/2}\right\} dt} + \nonumber
  \\
  &+& E_0 \int_{t_2}^{t_1}{q_h(t) \left[1 - \frac{\left<\rho r\right>_3}{\mu_3\left(E_0\right)}\right]^{1/2} dt} = \nonumber
  \\
  &=& E_0 N_0 \eta_h \eta_h^{-1} \tau_h^{-1} \left[ \int_{t_0}^{t_1}{\delta_h(t) \left[1 - \frac{\left<\rho r\right>_3}{\mu_3\left(E_0\right)}\right]^{1/2} dt} - \right. \nonumber
  \\
  &-& \left. \int_{t_0}^{t_2}{\delta_h(t) \left[1 - \frac{\left<\rho r\right>_3}{\mu_3\left(E_0\right)}\right]^{1/2} \left\{1 - \left(1 - \frac{\left<\rho r\right>_2}{\mu_2\left(E_1\right)}\right)^{1/2}\right\} dt} \right].
\end{eqnarray}
All energy that is not deposited in the regions 3 and 2, as was justified above, is deposited in the DT-fuel:
\begin{equation}\label{eq:022}
  \Delta E_1 = E_0 N_0 \tau_h^{-1} \int_{t_0}^{t_2}{\delta_h(t) \left[1 - \frac{\left<\rho r\right>_3}{\mu_3\left(E_0\right)}\right]^{1/2} \left(1 - \frac{\left<\rho r\right>_2}{\mu_2\left(E_1\right)}\right)^{1/2} dt}.
\end{equation}
Below the characteristics of fast electron deceleration and the energy transferred to the different parts of target calculated according to the above presented model are discussed. Table~\ref{tab:001} shows the time-integral characteristics of slowing down of fast electrons with the minimum and maximum energies of the range~(\ref{eq:001}) and with the most likely value of $E_0 = 50$~keV.
\begin{table}[!ht]
  \caption{The time moments $t_1$ and $t_2$, fractions of the number of heating fast electrons which hit into compressed ablator $\delta N_2 = N_2 / N_0 \eta_h$ and DT-layer $\delta N_1 = N_1 / N_0 \eta_h$, as well as, the fractions of their energy that heating fast electrons transfer to the evaporated ablator with overcritical density $\delta \Delta E_3 = \Delta E_3 / N_0 E_0 \eta_h$, compressed ablator $\delta \Delta E_2 = \Delta E_2 / N_0 E_0 \eta_h$ and DT-fuel $\delta \Delta E_1 = \Delta E_1 / N_0 E_0 \eta_h$ for different values of the initial energy of fast electrons from the range~(\ref{eq:001}).\label{tab:001}}
  \begin{indented}
    \item[]\begin{tabular}{@{}llllllll}
      \br
      $E_0$~(keV) & $t_1$~(ns) & $t_2$~(ns) & $\delta N_2$ & $\delta N_1$ & $\delta \Delta E_3$ & $\delta \Delta E_2$ & $\delta \Delta E_1$ \\
      \mr
      30 & 7.13 & --   & 0.11 & 0.0  & 0.98 & 0.02 & 0.0  \\
      50 & 8.88 & 8.44 & 0.88 & 0.78 & 0.41 & 0.18 & 0.41 \\
      70 & 10.0 & 9.83 & 1.0  & 0.99 & 0.18 & 0.08 & 0.74 \\
      \br
    \end{tabular}
  \end{indented}
\end{table}

\begin{figure}[!ht]
  \centering
  \includegraphics[width=0.6\textwidth]{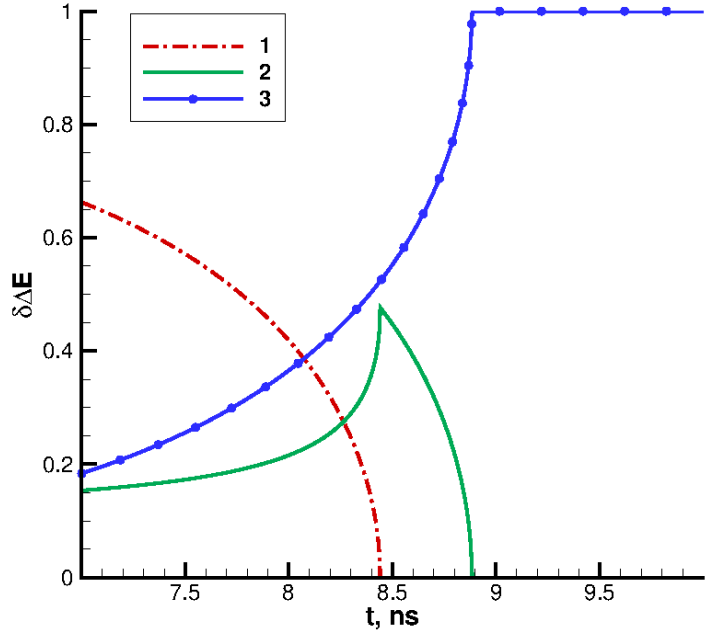}
  \caption{The time dependencies of the fraction of energy which is transferred by heating fast electrons with initial energy $E_0 = 50$~keV. Curve 1 corresponds to the DT-fuel, curve 2 -- to the layer of compressed ablator, curve 3 -- to the region of evaporated ablator with overcritical density.\label{fig:006}}
\end{figure}

\begin{figure}[!ht]
  \centering
  \includegraphics[width=0.6\textwidth]{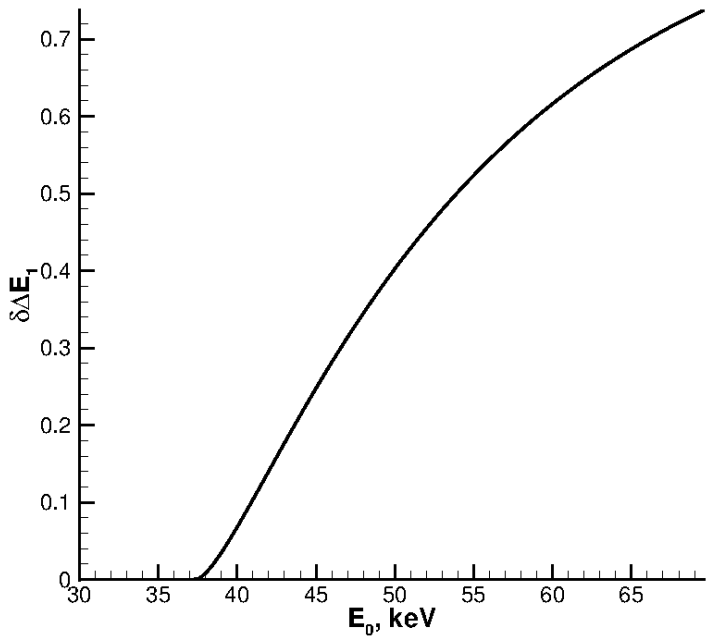}
  \caption{The dependence of the fraction of energy, which the heating fast electrons transfer to the DT-fuel, on the initial energy $E_0$.\label{fig:007}}
\end{figure}

Small fraction of heating fast electrons with initial energy of 30~keV hits into a compressed ablator, namely, only 11\% of their total number. The particles, got into compressed ablator, are generated during short time period from 7 to 7.13~ns, while the areal density of evaporated ablator does not exceed the mass range of fast electron with the initial energy $E_0 = 30$~keV. Almost all of their energy (98\%) leaves in the evaporated ablator, and only 2\% -- in the compressed ablator. Fast electrons do not hit into the DT-layer and do not heat it at all. However, due to the quadratic growth of fast electron mass range with increasing the energy $E_0$, the beginning of a significant warming up of DT-layer with increasing $E_0$ begins very quickly. Already at energy $E_0 = 50$~keV, 78\% of the heating fast electrons get into DT-layer, where they leave 41\% of their energy remaining after slowing down in the evaporated and non-evaporated parts of ablator. Particles reaching the DT-layer are generated during time period from 7 to 8.44~ns, while the total areal density of the evaporated and non-evaporated parts of the ablator does not exceed the mass range of the fast electron with initial energy $E_0 = 50$~keV. At energy of $E_0 = 70$~keV, almost all fast electrons reach DT-layer and transfer to it 74\% of their energy. On Fig.~\ref{fig:006} the time dependencies of the fractions of energy, which the heating fast electrons with initial energy $E_0 = 50$~keV transfer to the different parts of target, are shown.
The fraction of energy transferred to DT-fuel decreases with time due to the growth with time of the areal density of the evaporated and non-evaporated parts of ablator. As mentioned above, at 8.44~ns, this value becomes 0. The fraction of energy transferred to the compressed ablator increases with time, up to the time moment about 8.4~ns, until the areal density of the evaporated part of ablator (it grows with time) begins to exceed the areal density of the compressed part of ablator. After that, the fraction of energy transferred to the compressed ablator  begins to decrease with time, and the fraction of energy transferred to the evaporated ablator continues to grow. This growth continues until at the moment 8.88~ns when  energy transfer to the compressed ablator stops (see Table~\ref{tab:001}). After that, the energy is transferred only to the evaporated ablator. The dependence of the fraction of energy, which the heating fast electrons transfer to DT-fuel, on initial energy $E_0$ is shown on Fig.~\ref{fig:007}.
Thus, for the range of initial energy $30~\mbox{keV} \leq E_0 \leq 70~\mbox{keV}$, the opposite regimes of energy transfer from fast electrons to DT-fuel are realized: from almost no heating at 30~keV to the transfer of a significant fraction of energy, specifically 74\%. Another important conclusion is that for the considered range of initial energy, a large fraction of fast electrons energy is transferred to the evaporated part of ablator with overcritical density. It should be noted that in addition to the heating fast electrons, the electrons generated in a solid angle, which is formed by tangents to the $n_s$- and the $n_c$-surfaces, will be slowed down in this region of target. Additional energy transfer to the region of subsonic expansion contributes to an increase in the ablation pressure and, as a consequence, to an increase in the target implosion velocity. Thus, for the considered target, the transfer of a significant fraction of energy to the region of the evaporated part of ablator with overcritical density will partially compensate a negative effect of heating of the compressed shell.

For the first time, the contribution of energy transfer by fast electrons to  ablation pressure was discussed in~\cite{Guskov1983} as applied to the long-wavelength CO$_2$ laser interaction with matter. At present time it is paid much attention to the study of this effect in relation to the promising approach to ICF target ignition known as ``shock ignition''~\cite{Scherbakov1983,Betti2007}. In both cases, the effect is strong, since both cases correspond to a significant fraction of the laser energy contained in fast electrons -- up to 30\%. In the first case, this is due to the large laser wavelength, which is ten times longer than the wavelength of the fundamental harmonic of Nd laser. In the second case, this is due to the high intensity of a laser pulse, about $10^{16}$~W/cm$^2$, which is necessary for generation of an igniting shock wave. The effect of increasing ablation pressure due to energy transfer by fast electrons to dense plasma regions was experimentally observed in experiments~\cite{Guskov2004,Guskov2006} and in other ones~\cite{Theobald2008,Theobald2012,Guskov2014} devoted to study of shock ignition.

The influence of heating the target by fast electrons can be considered on the base of estimation of decreasing the final density of DT-fuel, which can be performed in the approximation of adiabatic compression law
\begin{equation}\label{eq:023}
  \frac{\rho_h}{\rho_0} \approx \left(\frac{T_0}{T_0 + \Delta T_h}\right)^{\gamma - 1},
\end{equation}
where $\gamma$  is specific heats ratio, $T_0$ is temperature of DT-fuel at the beginning of its deceleration $t = \tau_L = 10$~ns in the case of absence of fast electrons, $\rho_0$ and $\rho_h$ are the values of the final density of DT-plasma in the case of absence of fast electron generation and in the case of heating a DT-fuel by fast electrons, leading to an increment of temperature $\Delta T_h$; the density $\rho_0$ is about 600~g/cm$^3$, according to numerical simulation of the considered target without taking into account a fast electron generation.

The temperature increment $\Delta T_h$ can be approximately defined as
\begin{equation}\label{eq:024}
  \Delta T_h \approx \frac{\Delta E_1}{C_v M_s},
\end{equation}
where $C_v = 1.15 \cdot 10^{15}$~erg$\cdot$(g$\cdot$keV)$^{-1}$ is specific heat of a fully ionized DT-plasma.

\begin{figure}[!ht]
  \centering
  \includegraphics[width=0.6\textwidth]{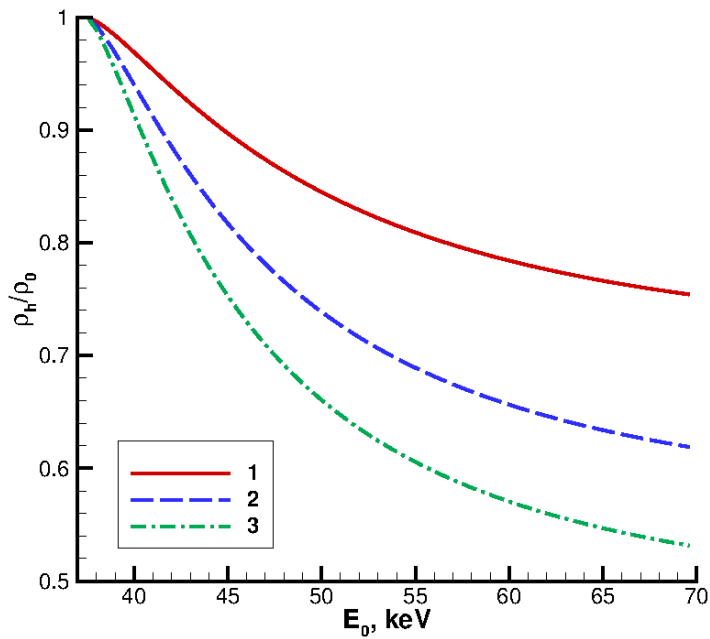}
  \caption{The dependence of the ratio $\rho_h / \rho_0$ on the initial energy of fast electrons for different values of the fraction of laser energy contained in fast electrons. Curve 1 corresponds to $\eta = 0.005$, curve 2 -- $\eta = 0.01$ and curve 3 -- $\eta = 0.015$.\label{fig:008}}
\end{figure}

Figure~\ref{fig:008} shows the dependencies of the ratio $\rho_h / \rho_0$ on the initial energy of fast electrons for three different values of the fraction of laser energy contained in fast electrons from the range~(\ref{eq:002}): $\eta = 0.005$, $\eta = 0.01$ and $\eta = 0.015$. The absolute values of the energy of fast electron emission for these cases are 14~kJ, 28~kJ and 42~kJ, respectively.
It can be expected that only in the case of $\eta = 0.005$, heating the target by fast electrons will not lead to an ignition failure for the considered range of probable values of the initial energy $E_0$, because even at $E_0 = 70$~keV, the ratio $\rho_h / \rho_0$ does not decrease below 0.75. In the cases $\eta = 0.01$ and, especially, $\eta = 0.015$, the range of the initial energy allowed for ignition is significantly narrowed: in the case of $\eta = 0.01$ the maximum acceptable energy decreases up to 52-54~keV, and in the case of $\eta = 0.015$ -- up to 42-46~keV. It is clear that these conclusions are of a predictive nature. They have to be confirmed by numerical calculations of a thermonuclear gain, taking into account the heating of target by fast electrons. We are planning to carry out such calculations in the near future using 1D hydrodynamic code with a kinetic module for calculating the energy transfer by fast electrons, which takes into account the ``wandering'' effect. It is very important to have experimental confirmation of this effect, indirect evidence of which is the isotropic character of the particle generation in the corona plasma with a substantially subcritical density. In this regard, it should be noted that the calculations using the model presented above in the case when all the fast electrons enter the compressed part of the target (the ``wandering'' effect is absent, $\delta_h = 1$) show that the ignition of the considered target becomes difficult throughout the considered range~(\ref{eq:002}) of the values of fraction of laser energy containing in fast electrons. The maximum value of the initial energy of fast electrons should not exceed 40-45~keV. In the absence of the ``wandering'' effect, at $\eta = 0.005$, the density ratio $\rho_h / \rho_0$ decreases down to 0.58 instead of 0.85 in the case, when the wandering effect is taken into account; at $\eta = 0.01$ this ratio decreases up to 0.42 instead of 0.74 and at $\eta = 0.015$ this ratio decreases up to 0.32 instead of 0.66.

\section{Conclusion\label{sec:005}}

The effect of ``wandering'' of isotropically laser-accelerated fast electrons, due to remoteness of its generation region from the surface of a compressed part of ICF target has a general nature. A significant part of the fast electrons produced in the low-density part of target's corona does not hit into the compressed part of the target, whose surface is rapidly reduced as a result of implosion. For the ICF target at the implosion velocity of 300-400~km/s this effect leads to the fact that the fraction of fast electrons that can transfer their energy to the compressed part of target (the fraction of heating fast electrons) turns out to be enough small. For the typical
target considered in the paper, corresponding to absorbed laser energy of 1.5~MJ, this fraction is only 12\% of the total number of generated fast electrons. As a result, the degree of negative influence of fast electron on target's compression and, as a consequence, on a thermonuclear gain is significantly reduced.

The character of energy transfer from the heating fast electrons to the compressed parts of target, which are non-evaporated ablator and DT-fuel, depends significantly on the initial energy of fast electrons. For the range of initial energy of $30~\mbox{keV} \leq E_0 \leq 70~\mbox{keV}$, which was chosen based on the results of recent experiments~\cite{Rosenberg2018}, opposite (from  the point of view of DT-fuel heating)  regimes of the energy transfer from fast electrons are realized. If the initial energy of the particles is close to the lower limit of this range, then the regimes of weak heating of the compressed part of the target are realized. At $E_0 \leq 37~\mbox{keV}$, fast electrons do not reach DT-fuel at all: at $29~\mbox{keV} \leq E_0 \leq 37~\mbox{keV}$, they are slowed down in the evaporated ablator with overcritical density and in the compressed part of ablator; at $E_0 \leq 29~\mbox{keV}$ -- only in the evaporated ablator with overcritical density. In the latter case, fast electrons will not adversely affect a compression of target. Moreover, the transfer of energy by fast electrons to the region of subsonic expansion of a matter with a density higher than critical one will contribute to an increase in the ablation pressure. Thus, generation of the fast electrons with an energy of about 30~keV will have a positive effect on compression. Fast electrons with energies close to the upper limit of the discussed range~(\ref{eq:001}) will transfer a significant part of their energy to DT-fuel. So, at the initial energy $E_0 = 70~\mbox{keV}$, the heating fast electrons transfer 74\% of their energy to DT-fuel. In this case, they transfer, respectively, 8 and 18\% of their energy to the compressed part of ablator and to the evaporated part of ablator with overcritical density. Note that the transfer of a significant fraction of fast electron  energy to the evaporated part of ablator with overcritical density is a general regularity for the considered range of initial energy of particles. For the electrons with initial energy $E_0 = 30~\mbox{keV}$ this fraction is 98\%, with $E_0 = 50~\mbox{keV}$ -- 41\% and with $E_0 = 70~\mbox{keV}$, as already mentioned -- 18\%. At $E_0 < 29~\mbox{keV}$, when fast electrons do not transfer their energy to the compressed part of target, the additional energy transfer by these particles to the supersonic region of a corona will only have a positive effect on a compression of target, at $E_0 > 29~\mbox{keV}$ it will partially compensate the negative effect of heating the compressed part of target.

Heating the compressed part of target by fast electrons leads to a decrease in the final density of DT-fuel, which naturally can lead to a decrease in a thermonuclear gain down to the ignition breakdown. Specifically, in the case of converting the 0.5\% of laser energy into fast electron emission with initial energy 50~keV, the final density of DT-fuel decreases by about of 1.2 times, in the case of $\eta = 0.01$ -- by about of 1.3 times and in the case of $\eta = 0.015$ -- by about of 1.5 times. In the absence of the ``wandering'' effect, the decrease in the final density turns out to be much larger: at $\eta = 0.005$ this decrease is 1.7, at $\eta = 0.01$ -- 2.4 and at $\eta = 0.015$ -- 3.1. Since the ``wandering'' effect can lead to a strong decrease in target preheating by fast electrons and preserving, thereby, the final density of DT plasma sufficient for ignition, experimental confirmation of this effect is important. Its indirect evidence may be an isotropic character of fast electron generation in the corona region with a substantially subcritical density. The found features of energy transfer by fast electrons such as the ``wandering'' effect and the transfer of a significant fraction of fast electron energy to the plasma of evaporated part of a ablator with overcritical density can have a significant effect on an implosion and, as a result, on a gain of direct-drive ICF target intended for ignition. We are planning to conduct studies of such an effect on the basis of numerical hydrodynamic calculations with a kinetic description of energy transfer by fast electrons in the near future.

\ack\label{sec:007}
The development of numerical algorithms and their incorporation into numerical codes were carried out with the financial support of the Russian Science Foundation under project No.~16-11-10174. The theoretical investigation of fast electrons kinetics was carried out with the financial support of Russian Foundation for Basic Research under the project No.~17-02-00059-a.

\section*{References}

\bibliographystyle{iopart-num}
\bibliography{refs}

\end{document}